\documentclass[article,12pt]{article}

\usepackage{amsthm}
\usepackage{subfigure}
\usepackage{graphicx}
\usepackage{amssymb,amstext,amsmath}
\usepackage{float}
\usepackage{bbm}
\usepackage{bm}
\usepackage{multirow}
\usepackage{makecell}
\usepackage{color}
\usepackage{appendix}
\usepackage{caption}
\usepackage{verbatim}
\usepackage{url}

\usepackage{tikz}
\usetikzlibrary{arrows}

\def\Item$#1${\item $\displaystyle#1$
\hfill\refstepcounter{equation}(\theequation)}

\newcommand\blfootnote[1]{%
  \begingroup
  \renewcommand\thefootnote{}\footnote{#1}%
  \addtocounter{footnote}{-1}%
  \endgroup
}

\UseRawInputEncoding

\begin{document}

\title{
An index of effective number of variables for uncertainty and reliability analysis in model selection problems
 }

\author{Luca Martino$^{\dagger}$, Eduardo Morgado$^{\dagger}$, Roberto San Mill{\'a}n Castillo$^{\dagger}$
       }


\date{}

\maketitle
\blfootnote{$^{\dagger}$ Universidad Rey Juan Carlos, Campus de Fuenlabrada, Madrid}

\begin{abstract}
An index of an effective number of variables (ENV) is introduced for model selection in nested models. This is the case, for instance, when we have to decide the order of a polynomial function or the number of bases in a nonlinear regression, choose the number of clusters in a clustering problem, or the number of features in a variable selection application (to name few examples). {It is inspired by the idea of the maximum area under the curve (AUC).}  The interpretation of the ENV index is identical to the effective sample size (ESS) indices  {concerning} a set of samples. 
The ENV index improves {drawbacks of} the elbow detectors described in the literature and introduces {different confidence measures of the proposed solution}. These novel measures can be also employed jointly with the use of different information criteria, such as the well-known AIC and BIC, {or any other model selection procedures}. Comparisons with classical and recent schemes are provided in different experiments involving real datasets. Related Matlab code is given.
\end{abstract}





\section{Introduction}
\label{sec:intro}
\begin{figure}[!ht]
\centerline{
          \includegraphics[width=0.65\columnwidth]{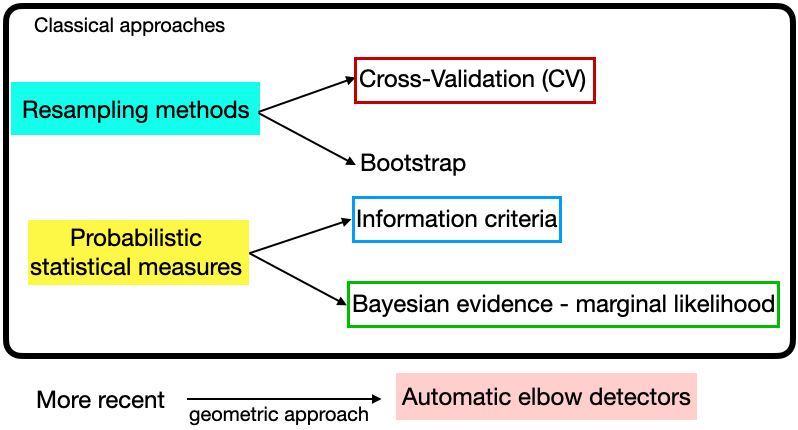}       
          }
   \caption{Classes of methods for model selection (standard ones and more recent approaches).}
   \label{fig:lNTRO}
\end{figure}

Model selection is absolutely a fundamental task of scientific inquiry \cite{Aho14,GUPTA2022,Hjort03,stoica2022monte}. It consists of selecting one model among many candidate models, given some observed data. We can distinguish three main frameworks in model selection. {The first scenario} is when completely different models are compared. The second setting is when several models defined by the same parametric family are considered (namely, these parameters of the model are tuned).  The third setting is related to the previous one but, in this scenario, the family contains models of different complexity since the number of parameters can grow (i.e., the dimension of the vector of parameter grows, building more complex models). This last case, also known as {\it nested models}, is what we address in this work. Examples of model selection in nested models are the order selection in polynomial regression or autoregressive schemes, variable selection,  clustering, and dimension reduction, just to name a few  \cite{COBOS2014,GKIOULEKAS2019,mukherjee2006nested,OurPaperSound,zhu2017bayesian}. { Other important applications in signal processing are the number of source signal estimation \cite{beheshti2018number} and the structured  parameter selection \cite{JANSEN2024104437}. 
}
\newline
In the selection of nested models, the decision is driven by the so-called bias-variance trade-off{\footnote{{A bias-variance trade-off could appear also in scenarios with non-nested models.}}}: we have {to choose} a compromise between (a) the model performance and (b) the model complexity.  Thus, the concept of selecting the best model is, in some sense, related to  {a proper mathematical definition} of a `good enough' model.  Hence, 
the issue is to properly describe in terms of equations  the colloquial expression  `good enough' \cite{bishop2006pattern}.
\newline
In the literature, there are two main families of methods for model selection, which are composed of three main sub-families, as depicted in Figure \ref{fig:lNTRO}.
 The first class is formed by {\it resampling methods}, where a main sub-family is given by the {\it bootstrap} and {\it cross-validation} (CV) techniques \cite{fong2020marginal,vehtari2017practical,Stoica04digital}. For simplicity, we focus here on CV. More specifically, CV is based on the splitting of the data in training and test sets. The training set {is used} to fit a model and the test set {is} to evaluate it. However, the proportion of data to use in training (and/or in testing) must be chosen by the user and can critically affect the results in terms of penalization of the model complexity. Moreover,  the splitting process should be repeated several times, and the performance can be averaged over the runs (that is computationally costly).

\noindent 
The second family is given by the so-called {\it probabilistic statistical measures}, which {employ different rules} for evaluating the different models considering directly the entire dataset (unlike in CV). This family is formed by two main sub-classes: the {\it information criteria} \cite{Ando11,COBOS2014,konishi2008information,VDLINDE05} and the {\it marginal likelihood} approach (a.k.a., Bayesian evidence) used in Bayesian inference \cite{Robert04,stoica2022monte,llorente2020review}.
Some famous information criteria are  the Akaike information criterion (AIC), which is based on the entropy maximization principle \cite{Spiegelhalter02}, and the Bayesian information criterion (BIC) which is also a bridge with  marginal likelihood approach, since BIC is derived as an approximation of the marginal likelihood  \cite{schwarz1978estimating}. Other very similar or completely equivalent schemes can be also found \cite{Foster94,Mallows73,RISSANEN78}. 
  \newline
 All the information criteria (IC) consider the sum of a performance measure and a penalty of the complexity. More specifically, they  employ  a linear penalization of the model complexity, where a parameter $\lambda$ represents a positive slope value of this penalty. They differ for the {\it slope} $\lambda$ of this linear penalization term \cite{mariani2015model} (see, for instance, Table \ref{TablaIC}). The choice of this slope is justified by different theoretical derivations, each one with several assumptions and approximations. Clearly, the results depend on this choice \cite{mariani2015model,SICpaper}. Some considerations regarding {the consistency} of IC can be found in {this} survey  \cite{Dziak20}.
The last approach is based on the marginal likelihood, {which is employed} in Bayesian inference for model selection purposes.  The model penalization in the marginal likelihood is induced by the choice of the prior densities \cite{SafePriorsLlorente,stoica2022monte}. 
  \newline
   \newline
 Therefore, in the three main approaches described above, (a) CV, (b) information criteria, and (c) marginal likelihood computation, we have always a degree of freedom (proportion of split data, slope $\lambda$, and choice of the prior densities) that affects the final results of the model selection.
For this reason, other recent approaches have been also investigated in the literature, based on geometric considerations: some of them propose {an automatic} detection of an ``elbow'' or ``knee-point'' in {a non-increasing curve describing a metric of performance of the model versus its complexity} \cite{AEDpaperNuestro,Elbow2,Elbow3,Elbow4}.  In \cite{AEDpaperNuestro}, the authors provide four different and equivalent geometric derivations showing:  (a) the elbow detectors in \cite{AEDpaperNuestro,Elbow2,Elbow3,Elbow4} are equivalent (providing the same result), and (b) this result can be obtained as an optimization of an information criterion with {a specific choice} of $\lambda$  (given in \cite{AEDpaperNuestro}), i.e., this geometric approach can be also expressed as an information criterion. On the other hand, another  recent information criterion, called spectral information criterion (SIC) \cite{SICpaper}, has been also introduced in the literature: this criterion uses all the possible values  of  $\lambda$, thus also contains the rest of IC as special cases, and returns also a confidence measure of the proposed solution.
\newline
\newline
In this work, we extend one of {the derivations} proposed in  \cite{AEDpaperNuestro} designing an index of {an} effective number of variables (ENV). Note that, {throughout} all the work, we use the words  {\it variables}, {\it components}, {\it features},   and/or {\it parameters} of a model as synonymous \cite{Elbow_paper,VarSelRev_paper}. The resulting index is inspired by the concept of maximum area-under-the-curve (AUC) in receiver operating characteristic (ROC) curves  \cite{bishop2006pattern,HanleyAUC82} and the Gini index, {widely used in economics} \cite{Lorenz05,Gini12,Gini13,Gini21}. Moreover, the underlying idea and interpretation  {are} exactly like the effective sample size (ESS) indices {concerning} a set of samples \cite{ESS1,ESS2}. {An important property of the ENV index is that} improves the elbow detectors removing the dependence on the maximum number of variables $K$ considered in the analysis \cite{AEDpaperNuestro,Elbow2,Elbow3,Elbow4}.
\newline
Moreover, we introduce measures of reliability and uncertainty of the proposed results, related to the ENV index, {similar} to the SIC outputs. They provide quantities {associated with} how `safe' is the solution, in {terms} of possible information lost by constructing a `too' parsimonious model.  
 It is important to remark that the novel  confidence measures can be employed also for different information criteria, such as AIC and BIC (to name a few), {or other feature importance and model selection schemes \cite{ShapleyValuesAndLOCO,GeneralizedVariableImportance23},} not only when {the elbow detectors are} applied.
In order to define these confidence measures, we also introduce some variable importance measures (see also the sensibility analysis in \cite{JorgePolonia}) that are in some way related to other relevant concepts already proposed in {the literature}, such as the Shapley values and feature importance ideas, {which are currently a hot research topic \cite{Shapley1,Shapley2}.  The fact that several information criteria and feature importance schemes have been introduced in the literature (depending on the specific context and application) is a clear signal that confidence measures for the results are required. }
\newline
The rest of {the work} is structured as follows. In Section \ref{sub:Ini_Curve}, we define the main notation and recall some background material. The ENV index is derived and analyzed in Section \ref{MAIN_SECT}. Additional properties of  the ENV index are given in Section \ref{Prop_Reliability}. Furthermore, in the same section, we introduce some confidence measures based on the derivation of {the} ENV index. In Section \ref{NEsect}, we show  different numerical experiments. We provide some final conclusions in Section \ref{COsect}. {Additional information is also contained in two appendices.}


 

{\color{red}




}

\section{Framework and main notation}\label{sub:Ini_Curve}
\subsection{The error curve $V(k)$ as {a} figure of merit}
\begin{figure}[!ht]
      \centerline{
      \subfigure[\label{Fig2Luca0}]{\includegraphics[width=0.35\columnwidth]{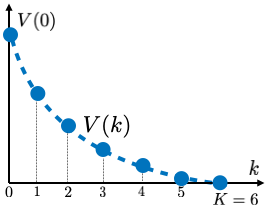}}
          \subfigure[\label{Fig2Luca}]{\includegraphics[width=0.35\columnwidth]{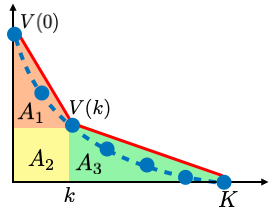}}
  }
   \caption{ {\bf (a)} Example of error function $V(k)$ where  $K=6$,   {\bf (b)} Construction with two straight lines and the areas $A_1$, $A_2${, and} $A_3$.}
   \label{fig:Curva_Inicial}
\end{figure}

In many applications in signal processing and machine learning, we desire to infer a vector of parameters ${\bm \theta}_k=[\theta_1,...,\theta_k]^{\top}$ of dimension $k$ given a data vector ${\bf y}=[y_1,...,y_N]^{\top}$.  A likelihood function $p({\bf y}| {\bm \theta}_k)$ is usually available, and often derived from a related physical model \cite{llorente2020review,Robert04}. 
 In many scenarios, also the dimension $k$ is unknown and must be estimated as well.  This is the case in numerous applications, for instance when $k$ can represent (a) the number of clusters in a clustering problem, (b) the order of a polynomial in a non-linear regression problem, (c) the number of selected variables in a feature selection problem, (d) the main number of dimensions in a dimension reduction problem, to name a few. 
In all these real-world application problems, a non-increasing {\it error function} (i.e., {a metric} that characterizes the performance of the system, such as a fitting measure),
$$
V(k): \mathbb{N} \rightarrow \mathbb{R}, \qquad k=0,1,2,...,K,
$$
  can be obtained.  Note that $k=0$ represents {the scenario of `no dependence', or `absence of model',} or the simplest possible model.
{Namely, $V(0)$ represents the value of the error function corresponding to  the simplest model, for instance, a constant model in a regression problem, or a single cluster (for all the data) in a clustering problem.} For instance, in polynomial regression, the value $V(0)$ could correspond to the variance of the data, i.e., the mean square error (MSE) error in prediction using a constant value $\theta_0$ (equal to the mean of the data) as {the} model. 
\newline 
 Generally, more complex models { - with more parameters, i.e., the vector ${\bm \theta}_k=[\theta_0,\theta_1,...,\theta_k]^{\top}$ has {a higher} dimension $k$ -} provide smaller errors in prediction/classification. We consider the most complex model having { $K$ parameters on top of the simplest model  the simplest case with only $\theta_0$}, i.e., ${\bm \theta}_K=[\theta_0,\theta_1,...,\theta_K]^{\top}$. {Without loss of generality,} we are considering an integer variable $k$ with increasing step of one unit. Clearly, more general assumptions could be {made}. In a variable/feature selection problem, we assume that the order of the variable inside the vector ${\bm \theta}_k$ is well-chosen, i.e., $V(k)$ is built after ranking the variables/features in a decreasing order of relevance \cite{OurPaperSound}.
 \newline
  \newline
 Examples of possible choices of $V(k)$ are the following:
 \begin{itemize}
 \item In the literature, when a likelihood function is given, a usual choice is  
$$
V(k)=-2\log(\ell_{\texttt{max}}), \quad \mbox{ where } \quad \ell_{\texttt{max}}=\max_{\bm{\theta}} p({\bf y}| \bm{\theta}_k),
$$ 
e.g., as in \cite{konishi2008information,schwarz1978estimating, Spiegelhalter02,llorente2020review}. 
\item $V(k)$ could be directly the MSE, and/or the mean absolute error (MAE) or transformations of them (as $\log \mbox{MSE}$, etc.). For instance,  in the linear and additive Gaussian noise case, it is possible to show that the choice $V(k)=-2\log(\ell_{\texttt{max}})$ is equivalent to $V(k)= N \log \mbox{MSE}$ where the MSE represents also an estimation of the noise power in the system (e.g., see \cite{WikiBIC}).
\item $V(k)$ can be any other error measure in classification or clustering, for instance. See Section \ref{VSwithOscar} for an example. 
\end{itemize}
{\bf Assumptions.} Generally, $V(k)$ should be a {\it non-increasing} error curve, i.e., for any pair of non-negative integers $n_1,n_2$ such that $n_2 > n_1$, then we have $V(n_2) \leq V(n_1)$.\footnote{This condition could be also relaxed. We keep it, for the sake of simplicity.} Indeed, $V(k)$ is a fitting term that decreases as the complexity of the model (given by the number $k$ of parameters) grows. Therefore, we have $V(0)\geq V(k), \forall k$. Note that $V(k)$ plays the same role  {as} the {so-called} Lorenz curve in the definition of the Gini index \cite{Lorenz05,Gini12}. {This general approach is similar to the framework used for defining the Shapley values \cite{ShapleyValuesAndLOCO,GeneralizedVariableImportance23}.} 
See Figure \ref{fig:Curva_Inicial} for a graphical example. 
\newline
 In some applications, the score function $V(k)$ should be also convex (as in Figure  \ref{fig:Curva_Inicial}), i.e., the differences $V(k+1)-V(k)$ will decrease as $k$ increases. This is the case of a variable selection problem if the variables have been ranked correctly.  However, this work does not require conditions regarding the concavity of $V(k)$. 
\newline
Finally,  just for the sake of simplicity and without loss of generality, we assume that $\min V(k)=V(K)=0$. Clearly, This condition can be always obtained with a simple subtraction. For instance, given a generic non-increasing function $V'(k)$, we can define  $V(k)=V'(k)-\min V'(k)=V'(k)-V'(K)$ so that $\min V(k)=0$.

\subsection{The universal automatic elbow detector}\label{sub:Calculo}

In this section, we briefly recall one of the derivations (all based on geometric arguments) of the universal automatic elbow detector (UAED) given in \cite{AEDpaperNuestro}. 
Here, we need to recall the derivation more connected to the AUC approach \cite{bishop2006pattern,HanleyAUC82}.  The underlying idea is to extract geometric information from the curve $V(k)$ {by} looking for a geometric ``elbow'' $k_e$ in order to determine the optimal number of components (denoted as $k_e\in \{0,1...,K\}$) to include in our model. 
\newline
We consider the construction of two straight lines passing through the points    
$(0,V(0))$, to $(k, V(k))$ and $(k, V(k))$, to $(K,0)$ as shown in Figure \ref{Fig2Luca} (where $k\in\{0,1,...,K\}$). Hence, we have piece-wise linear approximation of the curve $V(k)$ with these two straight lines. The goal is to minimize  
 the area under this approximation. More specifically, the total area under the approximation  is the sum of the two triangular areas  ($A_1$ and $A_3$) and the rectangular area ($A_2$) in Figure  \ref{Fig2Luca}. Namely, we have 
\begin{align*}
A_1 = \frac{k (V(0) - V(k))}{2},\  A_2 = k V(k), \ A_3 =\frac{(K - k) V(k)}{2}, 
\end{align*}
hence the optimal number of components $k_e$ (location of the ``elbow'') is defined as
\begin{align}\label{opt1}
k_e  = \arg\min_k\{A_1 + A_2 + A_3\}.
\end{align}
After some algebra, we arrive {at} the expression
\begin{align}
k_e  = \arg\min_k\left\{ V(k)+\frac{V(0)}{K} k  \right\}, \quad \mbox{for $k=1,...,K$}, \label{exp:N_opt}
\end{align} 
where we are assuming $V(0)\neq 0$ and $K\neq 0$.
\newline
\newline
{\bf Remark.} It is important to remark that, since $k$ belongs to a discrete and finite set, solving the optimization above is straightforward.
\newline
\newline
{\bf Remark.} If $V(k)$ is convex, $k_e$ is unique. Convexity of $V(k)$ is a sufficient but not necessary condition for the uniqueness of the solution.  
\newline
\newline
 {\bf Remark.}  If $V(k)$ is not convex, we can have several global minima. For instance, having $M$ different minima,
$ k^{*}_1, k^{*}_2,..., k^{*}_M$, the user can choose the best solution (within the $M$ possible one) according to some specific {requirements} depending on the specific application. Here, we suggest to pick the most conservative choice, i.e.,  $k_e=\max k^{*}_j$,  for $j=1,...,M$.
\newline
\newline
{\bf Relation with the information criteria.} The cost function employed by the UAED is 
\begin{align}\label{HereNewBIC}
C(k)&=V(k)+\frac{V(0)}{K} k,  \\
&=V(k)+\lambda k, \label{ICeq}
\end{align}
 where the slope of the complexity penalization term is $\lambda=\frac{V(0)}{K}$.
Therefore, this cost function has exactly the same form {as} the cost function used in the information criteria like BIC and AIC, i.e.,  with a linear penalization of the model complexity and selecting $\lambda=\frac{V(0)}{K}$. In  AIC and or BIC, we have $V(k)=-2\log \ell_{\texttt{max}}$. Therefore, when $V(k)=-2\log \ell_{\texttt{max}}$, the UAED can be interpreted as an information criterion with the particular choice of $\lambda=\frac{V(0)}{K}$. Table \ref{TablaIC} summarizes this information.

\begin{table}[!h]	
\caption{Information criteria in the literature, with the corresponding choices of $V(k)$ and $\lambda$. Note that $N$ {denotes} the number of data points and $\ell_{\max}$ is the maximum value reached by a likelihood function. The ENV index represents the scheme presented in this work.}\label{TablaIC}
\vspace{-0.3cm}
\begin{center}
\begin{tabular}{lcc} 
	\hline 
	{\bf Information criterion (IC)} & {\bf Choice of }$\lambda$  & {\bf$V(k)$}  \\ 
	\hline 
	Bayesian-Schwarz \cite{schwarz1978estimating}    & $\log N$  & $-2\log\ell_{\max}$ \\
	Akaike \cite{Spiegelhalter02}                    & $2$ & $-2\log\ell_{\max}$ \\
	Hannan-Quinn \cite{Hannan79}                     & $\log(\log(N))$  & $-2\log\ell_{\max}$ \\
     Universal Automatic Elbow Detector \cite{AEDpaperNuestro}  & $\frac{V(0)}{K}$ & any \\
 Spectral IC (SIC) \cite{SICpaper} & all & any \\ 
 \hline
  \hline
   Eff. num. of variables (ENV)  & --- & any \\
   \hline
\end{tabular}
\end{center}
\end{table}

\begin{figure*}[!ht]
      \centerline{
      \subfigure[\label{Fig2Luca2}]{\includegraphics[width=0.3\columnwidth]{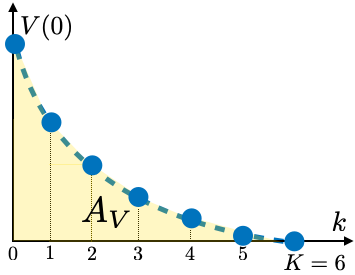}}
          \subfigure[\label{Fig2Luca3}]{\includegraphics[width=0.3\columnwidth]{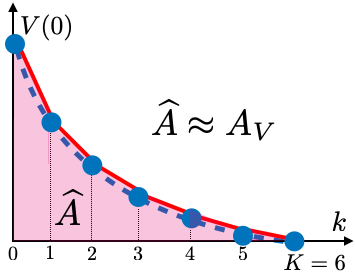}}
     }

   \caption{\footnotesize {\bf (a)} We can consider that $V(k)$ is like a sampled curve obtained from sampling - in a signal processing sense - a continuous function $V(x)$ where $x\in \mathbb{R}$ (shown in dashed line) is an auxiliary continuous variable. The  continuous function $V(x)$ possibly {does not} exist {and} can be just a theoretical tool to define the area $A_V$. {\bf (b)} In any case, we have {access} to $V(k)$, $k\in \mathbb{N}$, {which} allows {us} to obtain the approximation $\widehat{A} \approx A_V$. }
   \label{fig:AreaA}
\end{figure*}

\begin{figure*}[!ht]
          \centerline{
            \subfigure[\label{Fig2Luca4}]{\includegraphics[width=0.3\columnwidth]{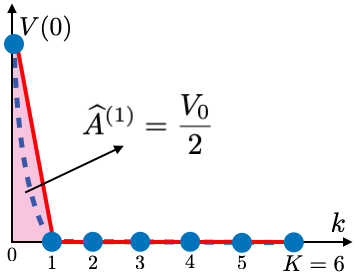}}
            \subfigure[\label{Fig2Luca5}]{\includegraphics[width=0.3\columnwidth]{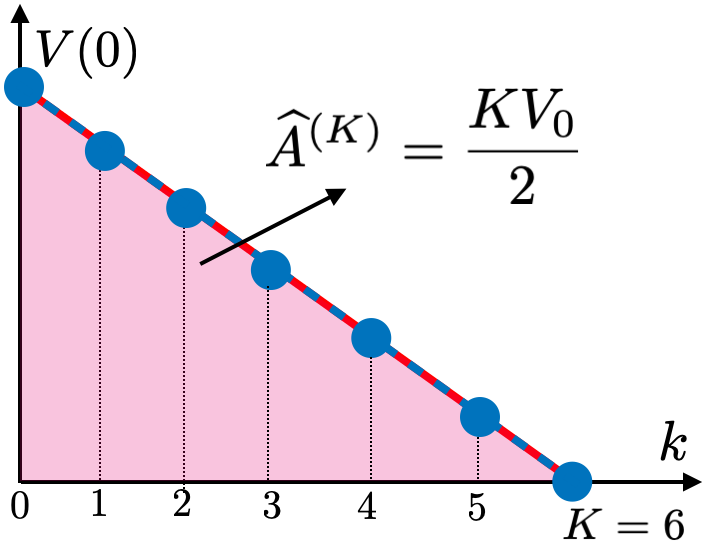}}
             \subfigure[\label{Fig2Luca6}]{\includegraphics[width=0.3\columnwidth]{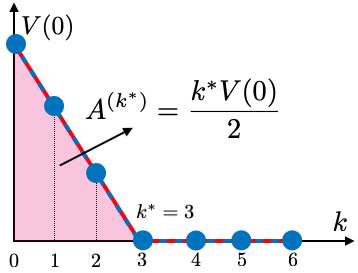}}
  }
   \caption{\footnotesize Special ideal cases.  {\bf (a)} $V(k)$ reaches zero already at $k=1$. Only the first component is relevant, hence the optimal choice $k^*=k_e=1$. {\bf (b)}  $V(k)$ is {a} straight line connecting {the points} $(0,V(0))$ and $(K,0)$. All variables contribute in the same way to the decay $V(k)$, hence the optimal choice $k^*=k_e=K$ (in figure $K=6$). {\bf (c)} The function $V(k)$ is a straight line passing through the points $(0,V(0))$ and $(k^*, 0)$, that is $V(k)=V(0)-\frac{V(0)}{k^*}k$, so that $V(k^*)=0$ at some point $k^*<K$. Clearly, the point $k^*=k_e=3$ is an optimal  choice:  the first 3 variables have the same contribution to the decay $V(k)$ and completely explain the drop.   }
   \label{fig:AreaA_specialCases}
\end{figure*}

\section{An index of the effective number of variables (ENV)}\label{MAIN_SECT}

The elbow detectors provide very good performance in several different scenarios, as shown in \cite{AEDpaperNuestro}. This {proves} the strength of the geometric approach compared to other information criteria proposed in the literature. 
\newline
However, all the elbow detectors presented in the literature \cite{AEDpaperNuestro,Elbow2,Elbow3,Elbow4} {have} a dependence on the maximum of variables analyzed, i.e., $K$. This is due to the slope of the linear complexity penalty  {being} $\lambda=\frac{V(0)}{K}$.  See also Section \ref{NumExample1}, for a numerical example. In some applications, the number $K$ is {the 
maximum} value strictly defined by a limitation of the system/model. In other frameworks, $K$ could be increased: if a new possible component is added even with a small impact on the decrease of error function $V(k)$, since the slope $\lambda=\frac{V(0)}{K}$ becomes smaller, the elbow detectors could suggest a bigger value as the optimal number of components. On the other hand, if the model selection analysis is performed {by} reducing the possible total number of components in advance, the elbow detectors would suggest a smaller value as the optimal number of variables.\footnote{Clearly, there is also a dependence of $V(0)$ and, more generally, the result depends on the choice of error curve $V(k)$, as well. } 
 Moreover, in several applications, it is also important to obtain a confidence measure for the model selection, jointly with the elbow detection. In the next sections, we try to address these issues.

\subsection{Derivation of the ENV index}

In this section, we extend the geometric approach employed {by} the UAED designing a new index that {\bf (a)} reduces the dependence on $K$ and {\bf (b)} helps us to provide confidence measures with respect to the chosen elbow. 
 The main idea is formed by two parts: {\bf (Part-1)} to obtain a better approximation $\widehat{A}$ (w.r.t. the  derivation of the UAED) of the area under the curve $V(k)$, denoted as $A_V$ and shown {in} Figure \ref{Fig2Luca2}, {\bf (Part-2)} then we have to normalize the obtained value $\widehat{A}$  (considering some ideal scenario). We provide more details below. 
\newline
\newline
{\bf (Part-1)} {Looking Figure  \ref{fig:AreaA},  we can observe that} a better approximation of the area $A_V$ under the function $V(k)$ can be easily obtained as sum of $K$ trapezoidal pieces, {as shown in Figure \ref{Fig2Luca3} with red lines}:
\begin{align}\label{mostImp0}
\widehat{A}&=\sum_{k=0}^{K-1}\frac{V(k)+V(k+1)}{2},  \nonumber  \\
&= \frac{V(0)}{2}+ \frac{V(1)}{2} + \frac{V(1)}{2}+ \frac{V(2)}{2}+...+ \frac{V(K-1)}{2}+ \frac{V(K)}{2},  \nonumber  \\
&=\frac{V(0)+V(K)}{2}+\sum_{k=1}^{K-1} V(k), \nonumber \\
&=\frac{V(0)}{2}+\sum_{k=1}^{K-1} V(k), 
\end{align} 
where, in the last equality, we have used the assumption that $V(K)=0$ and the step of increase on axis $k$ is $1$ (without loss of generality). A {graphical} example is depicted in Figure \ref{Fig2Luca3}.
\newline
\newline
{\bf (Part-2)}  In order to design a {\it normalized} index effective of the number of components, we need to describe an ideal scenario: let us assume that all entire drop of $V(k)$ is already reached in the first step {with a linear decay}, i.e., we have already $V(1)=0$ and by assumption the rest of values are also zero, $V(1)=V(2)=...V(K)=0$, as shown in Figure \ref{Fig2Luca4}. In this case, the first variable/component is relevant whereas the rest of {the} variables do not give any contribution to the decay of $V(k)$.  Therefore, the correct decision as effective {as} the number of components is $k^*=k_e=1$.\footnote{If we suppose a continuous function that sampled at each step $1$ generated the curve $V(k)$, the result $k^*=k_e=1$ is optimal for us, even if the function decays faster than a linear decrease,  but only because we have {\it not}  access to sampled points  between 0 and 1 (i.e., we have no information about the faster decay between 0 and 1); see Figure \ref{Fig2Luca4}. Having more information, i.e., more points between  0 and 1 (hence an increase in $k$ smaller than 1), then the optimal result would be $k^*<1$ (if we have a decay as the dashed line in Figure \ref{Fig2Luca4}).} The approximated area under the curve $V(k)$ in this case is $\widehat{A}^{(1)}=\frac{V(0)}{2}$. 
Thus, we define the index of {an} effective number of variables (ENV) as:
\begin{gather}\label{mostImp}
I_\texttt{ENV}=\left\{
\begin{split}
&0,  \qquad \qquad \quad \quad \mbox{ when }   V(0)= 0, \\
&\frac{\widehat{A}}{\widehat{A}^{(1)}}=\frac{2}{V(0)}\widehat{A}, \mbox{ when }  V(0)\neq 0.
\end{split} \right. 
\end{gather} 
Note that, when $V(0)\neq 0$, we can then write
\begin{align}\label{ENVindexEq2}
I_\texttt{ENV}=1+2\sum_{k=1}^{K-1} \frac{V(k)}{V(0)}, \qquad (\mbox{for $V(0)\neq 0$}). 
\end{align} 

\subsection{Behavior of $I_\texttt{ENV}$ in ideal cases}

In this section, we check the behavior of the ENV index in different extreme and ideal cases:

\begin{itemize}
\item In the ideal scenario above shown in Figure \ref{Fig2Luca4}, i.e., when $V(0)\neq 0$ but we have already $V(1)=0$ and $V(1)=V(2)=...V(K)=0$,  the correct decision is $k^*=1$. Hence, note also that $\widehat{A}=\widehat{A}^{(1)}$.  In this case, from \eqref{mostImp} or from \eqref{ENVindexEq2} we get $I_\texttt{ENV}=1$, as desired. 
\item Another ideal scenario is when $V(k)$ is a linear straight line connecting the points $(0, V(0))$ and $(K, V(K))$, as shown in Figure \ref{Fig2Luca5}. In this situation, all the variables contribute in the same way {as} the decay of $V(k)$ (i.e., each variable has the same influence on the error decrease), so that the correct decision is $k^*=K$ (all the components are relevant, {with the same importance}). The area under the curve $V(k)$ in this case is $\widehat{A}^{(K)}=\frac{K V(0)}{2}$. In this case, the equality $\widehat{A}^{(K)}=A_V$ also holds. With Eq. \eqref{mostImp}, we also obtain $\widehat{A}=\frac{K V(0)}{2}$ therefore the ENV index is $ I_\texttt{ENV}=\frac{2}{V(0)}\frac{K V(0)}{2}=K$, exactly as expected.
\item Another extreme case is when all the components are independent {of} the output $y$. In this situation, theoretically $V(k)$ should be a constant {(i.e., there is not any drop)}, $V(k)=V(0)$ for all $k$, namely $V(0)=V(1)=....V(K)=0$ due to our assumption $V(K)=0$, without losing any generality. Hence, the correct decision is $k^*=0$  and the area under the function $V(k)$ is $\widehat{A}^{(0)}=0$.  {In this} case,  $\widehat{A}=0$ so that $I_\texttt{ENV}=0$, again as desired.
\item More generally, consider that at some integer $k^*$ we have $V(k^*)=0$, and the $V(k)$ is a straight line passing through the points $(0,V(0))$ and $(k^*, 0)$, that is $V(k)=V(0)-\frac{V(0)}{k^*}k$.  Then,
\begin{align*}
I_\texttt{ENV}&=1+2\sum_{k=1}^{k^*} \frac{V(0)-\frac{V(0)}{k^*}k}{V(0)}, \\
&=1+2\sum_{k=1}^{k^*} \left(1-\frac{k}{k^*}\right), \\
&=1+2 k^*-\frac{k^*(k^*+1)}{k^*},  \\
&=1+2 k^*-(k^*+1), \\
&=k^*. 
\end{align*} 
Exactly as expected and desired. This can be also easily obtained {by looking at} FigureÊ \ref{Fig2Luca6}: indeed, in this scenario we can write   $\widehat{A}=\frac{k^* V(0)}{2}$ and the ENV index will be $ I_\texttt{ENV}=\frac{2}{V(0)}\frac{k^* V(0)}{2}=k^*$.
\end{itemize}
  Hence, we have an index such that
\begin{align} 
0\leq I_\texttt{ENV} \leq K.
\end{align}
Some additional properties and observations are given below and {in Appendix \ref{APP1}}.

\section{Interpreting and using the ENV index}\label{Prop_Reliability}
\noindent
In this section, we introduce {some properties} of the ENV index, and {their} effects are shown in Section \ref{NumExample1}. An interesting behavior of $I_\texttt{ENV}$ in ideal scenarios is described below and depicted in Figure \ref{fig:Picasso_comeme_el_casso}. More generally, we explain {how to interpret and use}  $I_\texttt{ENV}$.

\subsection{First considerations} \label{FC_SECT}

{\bf Property 1.} Let us consider a generic non-increasing function $V'(k)$, with $k=0,...,K${, and} assume to have {a} horizontal asymptote, i.e.,  $\lim_{K\rightarrow \infty} V'(k)=C$.  We can always define $V(k)=V'(k)-\min V'(k)$ so that $V(K)=0$ for each possible $K$.  Note that the ENV index in Eq. \eqref{ENVindexEq2} depends on $K$ (hence we use the notation here $I_\texttt{ENV}=I_\texttt{ENV}(K)$) but, when $K\rightarrow \infty$, we have that {\it stability property}, i.e.,
\begin{align}
\lim_{K\rightarrow \infty} I_\texttt{ENV}(K)=\bar{I}_\texttt{ENV},
\end{align} 
i.e., the ENV index converges to a stable fixed value $\bar{I}_\texttt{ENV}$, that represents a geometrical feature of the curve $V(k)$. This {is because} adding infinitesimal portions of {the} area to $A_V$ virtually does not change the values of $A_V$ itself and $\widehat{A}$; see Figs. \ref{Fig2Luca2} and \ref{Fig2Luca3}.  Generally, this is not the behavior of the elbow detectors \cite{AEDpaperNuestro,Elbow2,Elbow3,Elbow4} in (non-ideal) real scenarios. See the numerical example in Section \ref{NumExample1}. 
\newline
\newline
{\bf Property 2.}  Moreover, another property is that 
\begin{align}
 I_\texttt{ENV}\geq 1, \quad  \mbox{for  $V(0)\neq0$},
\end{align}
as we can observe clearly from Eq. \eqref{ENVindexEq2}.
This is due to the fact that $k$ is a discrete variable with increasing step of $1$, i.e., $k=0,1,...,K$, and if $V(0)=0$ we have $V(k)=0$ for all $k$, by assumption.\footnote{We have access only to a ``sampled'' curve $V(k)$ {(sampled in the sense of sampling a continuous signal  to obtain a discrete signal).} We do not know $V(x)$ where $x\in\mathbb{R}$ that possibly defines a theoretical area $A_V$, as shown in Figure \ref{fig:AreaA}. If, in a specific problem, we have {access to $V(x)$ then} $I_\texttt{ENV}$ could take {any} intermediate values also between 0 and 1.  } 
\newline
\newline
{ {\bf Property 3.} $V(0)=0$ implies that $I_\texttt{ENV}=0$,} as also shown in Eq. \eqref{mostImp}. This  {is because} if $V(0)=0$ we have $V(k)=0$ for all $k$, by assumption. {See also Appendix \ref{APP1}.}
\newline
\newline
{ {\bf Property 4.}  $I_\texttt{ENV}$ is invariant to linear scaling of the curve $V(k)$. Namely, a different curve $V'(k)=a V(k)$ with $a>0$ has the same value $I_\texttt{ENV}$ of $V(k)$. Indeed, we have:
\begin{align*}
I_\texttt{ENV}' &=1+2\sum_{k=1}^{K-1} \frac{V'(k)}{V'(0)}=1+2\sum_{k=1}^{K-1} \frac{aV(k)}{aV(0)}=1+2\sum_{k=1}^{K-1} \frac{V(k)}{V(0)}=I_\texttt{ENV}. 
\end{align*}}
{\bf Choosing the number of components.} {The derivation of the ENV index is based on extending the UAED derivation, providing a better approximation of the area under the curve $V(k)$. Thus, as} an elbow detector,  the ENV index can be employed for choosing the number of components  in a model selection problem just {by} defining
 \begin{align} 
 k^*= \lfloor I_\texttt{ENV} \rceil, 
\end{align}
where $\lfloor a\rceil$ represents the rounding operation of replacing an arbitrary real number $a$ by its nearest integer. 
\newline
\newline
{ {\bf Interpreting $I_\texttt{ENV}$.} Figure \ref{fig:Picasso_comeme_el_casso} shows 4 ideal cases of $V(k)$ with a blue solid line. In all cases, the curve $V(k)$ is composed {of} two straight lines yielding an elbow at $k_e=14$. A well-designed elbow detector as in \cite{AEDpaperNuestro,Elbow2}, always picks $k_e=14$ as a solution, since this is the position of the elbow (hence this is the correct result for an elbow detector). The ENV index $I_\texttt{ENV}$ is shown with red triangles: it is able to discriminate the 4 different scenarios. {The} effective number of variables is exactly $k_e=14$ in the lowest line and becomes greater and greater as $V(k)$ becomes closer and closer to the red line (that represents the ideal scenario where all components are equally relevant).

 }
\subsection{Confidence measures for the decision}\label{SectReliab} 

\begin{figure}[!ht]
      \centerline{
    \includegraphics[width=0.5\columnwidth]{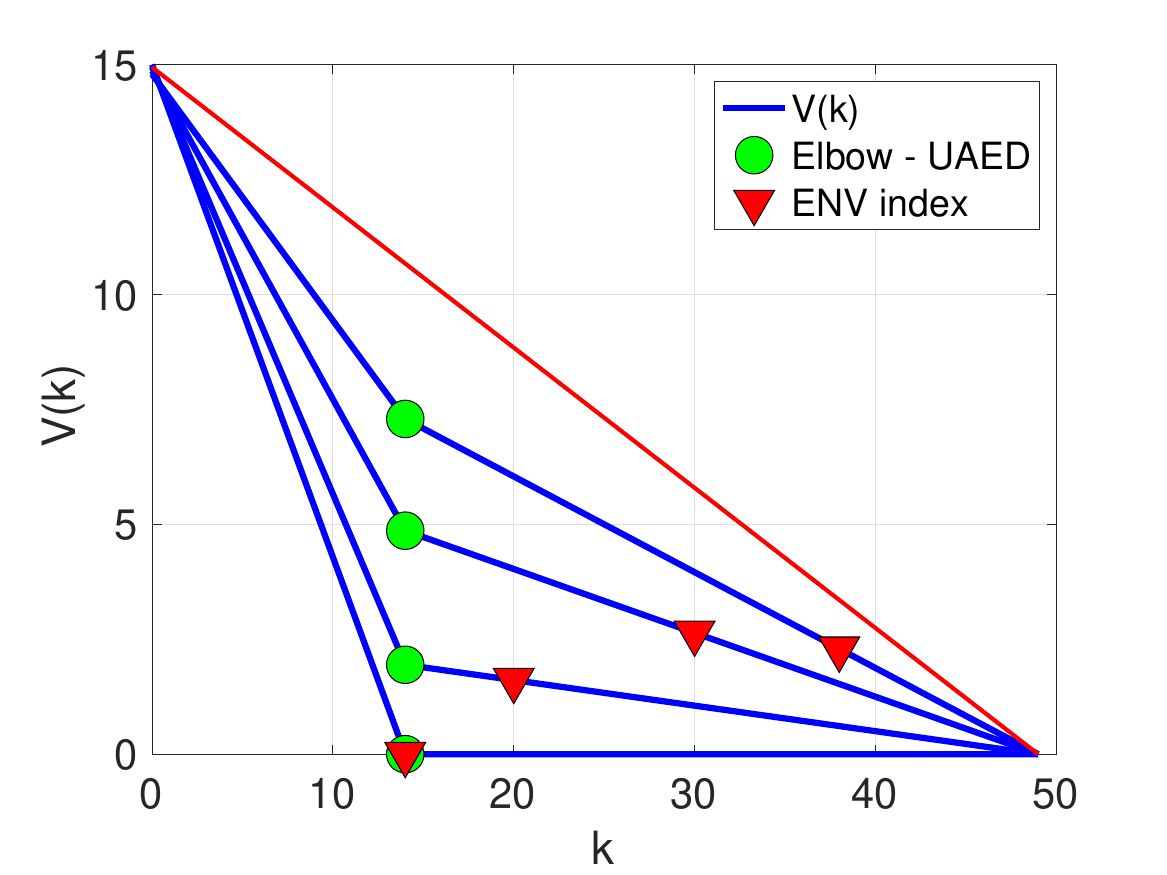}
  }
   \caption{Ideal cases of {four} curves $V(k)$ (blue solid lines) where an ``elbow'' is well-defined at $k_e=14$. This is exactly the result that we obtain with the application of an elbow detector \cite{AEDpaperNuestro,Elbow2,Elbow3,Elbow4} (green circles), whereas the results provided by the ENV index are shown by red triangles. Note that, in all cases, $I_\texttt{ENV}> k_e$.}
   \label{fig:Picasso_comeme_el_casso}
\end{figure}

{ The ENV index can be employed to detect the quality of a decision provided by an elbow detector, an information criterion, or another model selection procedure. For instance, as discussed above, the ENV index $I_\texttt{ENV}$ is able to discriminate the 4 different scenarios in Figure \ref{fig:Picasso_comeme_el_casso}, and determine the  confidence in the choice of the model with $k_e=14$ parameters. In the lowest curve $V(k)$, the choice $k_e=14$  is clearly an optimal point, since $V(k)$ already reaches zero at $k=14$. However, the confidence in this choice decreases as the blue line becomes closer and closer to the red line. The value $I_\texttt{ENV} $ (depicted in red triangles) becomes more and more distant to the elbow position $k_e$.
\newline
The goal is to design some mathematical tools/measures for evaluating the robustness and degree of certainty of the decision (i.e., the model choice). Before introducing some confidence measures that rely on $I_\texttt{ENV}$, we provide a definition of {\it variable importance} based on the same geometric considerations behind the ENV index. 
 }
\newline
\newline
{\bf Variable importance.} The underlying idea behind the UAED and the ENV index is to associate a measure of the importance of each variable/component, defined as a value proportional to its contribution to the decay of $V(k)$, i.e.,
$$
w_k=V(k-1)-V(k),  \quad k=1,...,K.
$$
Hence, $w_k$ represents the importance of {the} $k$-th component.
We can normalize these weights arriving {at} the following result: 
$$
 \bar{w}_k=\frac{w_k}{\sum_{i=1}^{K-1} w_i}=\frac{V(k-1)-V(k)}{V(0)},
 $$
 where we have used that
 $$
 \sum_{i=1}^{K-1} w_i=V(0)-V(K)=V(0),
 $$
 since $V(K)=0$ by assumption.  In the case of equally {important} variables, given when $V(k)$ is a unique straight line as in Fig. \ref{Fig2Luca5} or the red line in Fig.  \ref{fig:Picasso_comeme_el_casso},  we have 
\begin{align*}
&V(k)=V(0)-\frac{V(0)}{K}k, \\
&w_k=V(k-1)-V(k)=-\frac{V(0)}{K}(k-1)+\frac{V(0)}{K}k=\frac{V(0)}{K}, 
\end{align*} 
that is {the} same value for all $k$ (as expected) and, as {a} consequence, $\bar{w}_k=\frac{1}{K}$ for all $k$. This definition of variable importance {has} been already applied with success in \cite{JorgePolonia}.
\newline
 {\bf Cumulative importance (CI).} We can compute the importance accumulated using the first (ranked) $k$ components,
 $$
\mbox{CI}(k)= \sum_{i=1}^{k}  \bar{w}_i=\frac{ \sum_{i=1}^{k} w_i}{V(0)}=\frac{V(0)-V(k)}{V(0)}=1-\frac{V(k)}{V(0)}
 $$ 
 for $k=0,...,K$. Note that $0\leq \mbox{CI}(k)\leq 1$, $\mbox{CI}(0)=0$ and $\mbox{CI}(K)=1$. If {the zero value} is reached for a $k^*\leq K$, i.e., $V(k^*)=0$, we have $\mbox{CI}(k^*)=1$. This cumulative sum of normalized weights can play the same roles {as} the cumulative sum of normalized weights in SIC (i.e., as a reliability/safety measure), although the computation of the weights follows a completely different procedure \cite{SICpaper}.  In the case of equally {important} variables we have $\mbox{CI}(k)=\frac{k}{K}$. 
  The CI can be considered an accuracy measure: closer to 1 we have more accuracy, i.e., we have safer and more reliable decisions. Hence, as a consequence, $1-\mbox{CI}(k)$ is an uncertainty measure (the amount of {\it lost importance}, lost choosing a model with only the first $k$ components), and we can define also the cumulative uncertainty as
\begin{align}
\mbox{CU}(k)=1-\mbox{CI}(k)=\frac{V(k)}{V(0)}.
\end{align}
{ Note that CU resembles the definition of the so-called Sobol's indices \cite{Sobol93,klein2024Sobol}.}
    We can rewrite the ENV index as {the} sum of the cumulative uncertainties,
\begin{align}
I_\texttt{ENV}=1+2\sum_{k=1}^{K-1} \mbox{CU}(k), \quad \mbox{for $V(0)\neq 0$.}
\end{align}
In Figure \ref{fig:Picasso_comeme_el_casso}, from the bottom to the upper curve, the strengths of the $4$ elbows at $k_e=14$ are $\mbox{CI}(k_e)=1$, 0.87, 0.67, and 0.50, respectively.  
\newline
{\bf Reliability of the decision.} Other indicators of the reliability of the decision can be designed. 
Indeed, the value $I_\texttt{ENV}$ provides the {\it effective number} of variables/components in our model selection problem. Clearly, the decision {to use} $k<I_\texttt{ENV}$ variables is {less  safe (in terms of loss of information and performance)} instead of using a model with $k>I_\texttt{ENV}$. In this sense, a suitable indicator for the reliability of the decision could be defined as
\begin{align}
R_D=\min\left[1,\frac{ k_e }{I_\texttt{ENV}}\right].
\end{align}
Namely, any elbow position $k_e$ such that $k_e-I_\texttt{ENV} > 0$ can be considered in a safe region (since we are using  more variables than the effective number of variables), i.e., according to the ENV index, more parsimonious models could be chosen. { An alternative to these confidence measures has been introduced in \cite{SICpaper}; see Appendix \ref{APP2} for some details.}

\section{Numerical experiments}\label{NEsect}
In this section, we test the ENV index in different frameworks, including experiments with artificial data (the first one) and two real datasets (the last two experiments). We will see  that $I_\texttt{ENV}$ provides very good performance and presents more robustness with respect to other alternatives in the literature and we show the behavior of the confidence measures proposed above.\footnote{The Matlab code and datasets of the experiments are available at \url{http://www.lucamartino.altervista.org/PUBLIC_ENV_CODE.zip}.} 

\subsection{Synthetic experiment where $V(k)$ is an analytic function}
\label{NumExample1}
The goal of this example is to show the behaviors of the elbow detectors \cite{AEDpaperNuestro,Elbow2,Elbow3,Elbow4} and the ENV index in a synthetic framework but that is not an ideal scenario (as in Figure \ref{fig:Picasso_comeme_el_casso} where an elbow is well-defined). We consider the function
$$
V'(k)=e^{-0.1 k}, \quad k=0,1,2...,K.
$$
We consider different {values} of $K$. For each possible value of $K\in \{20, 50,500,5000\}$, we define 
$V(k)=V'(k)-\min V'(k)=e^{-0.1 k}-e^{-0.1 K}$, so that $V(K)=0$. We apply the  elbow detectors \cite{AEDpaperNuestro,Elbow2,Elbow3,Elbow4} and the ENV index obtaining the following results shown in Table \ref{Table0}.

\begin{table}[!h]
\caption{Results in the synthetic experiment. } \label{Table0}
\begin{center}
\begin{tabular}{|c|c|c|c|c|}
\hline
Methods & $K=20$  & $K=50$  & $K=500$  & $K=5000$   \\
\hline
\hline
 Elbow detectors  &  8 & 16 & 39  & 62 \\ 
 ENV index   &  13.756   & 19.338 & 20.016  & 20.016 \\
 \hline
 \hline
CI   & 0.58  &  0.78  & 0.98  & 0.99 \\
 $R_D$  & 0.58  & 0.83 &  1 & 1 \\
 \hline
\end{tabular}
\end{center}
\end{table}

Hence, the ENV index converges to the value $\bar{I}_\texttt{ENV} =20.016$. This stable value of the ENV index  is an intrinsic geometric property of the analyzed curve $V(k)$.  Moreover, we can observe that both confidence indices are quite smaller than 1 when the elbow is smaller than $\bar{I}_\texttt{ENV}=20.016$, and {both measures} approach $1$ as the elbow position becomes greater and greater, as expected. {We can also see} that the index of reliability of the decision $R_D$ is more optimistic than the cumulative importance (CI).

\subsection{Variable selection in a regression problem with real data}\label{Ex2Sound}
In a regression problem, we observe a dataset of $N$ pairs $\left\{\mathbf{x}_n, y_n\right\}_{n=1}^N$, where each input vector $\mathbf{x}_n=\left[x_{n, 1}, \ldots, x_{n, K}\right]$ is formed by $K$ variables, and the outputs $y_n$ 's are scalar values \cite{VarSelRev_paper}. We consider the case that $K \leq N$ and assume a linear observation model,
\begin{align}\label{ModelExpSoundScape}
y_n=\theta_0+\theta_1 x_{n, 1}+\theta_2 x_{n, 2}+\ldots \theta_K x_{n, K}+\epsilon_n,
\end{align}
where $\epsilon_n$ is a Gaussian noise with zero mean and variance $\sigma_\epsilon^2$, i.e., $\epsilon_n \sim \mathcal{N}\left(\epsilon | 0,\sigma_\epsilon^2 \right)$. More specifically, in this real dataset \cite{OurPaperSound,san2024variable,fan2017emo}, there are $K=122$ features and $N=1214$ number of data points ${\bf x}_i$. The dataset presents two outputs: "arousal" and "valence". In this section, we focus on the first output in the dataset  (``arousal'').
In this experiment, we can set $V(k)=-2 \log \left(\ell_{\max }\right)$ with $\ell_{\max }=\max _\theta p\left(\mathbf{y} | \boldsymbol{\theta}_k\right)$ with $k \leq K$, after ranking the 122 variables (see \cite{OurPaperSound}), where the likelihood function $p\left(\mathbf{y} | \boldsymbol{\theta}_k\right)$ is induced by the Eq. \eqref{ModelExpSoundScape}. Thus, here we can compare with other information criteria given in the literature, shown in Table \ref{TablaIC} and a  standard method based on the computation of p-values \cite{efroymson1960multiple}-
\cite{hocking1976biometrics}.  The ENV index returns a value of 12.74.
Thus, {the results provided} by each method are given in Table 
\ref{Table1}.

\begin{table}[!h]
\caption{Results in the variable selection example, in a regression problem with real data. } \label{Table1}
\begin{center}
\begin{tabular}{|c|c|c|c|c|c|c|c|c|}
\hline
{\bf Scheme} & p-value & AIC & BIC & HQIC & UAED & SIC-95 & SIC-99  &{\bf ENV} \\
\hline
$k_e$ &  71 & 44 & 17 &  41 &  11 & 7 & 17  &  {\bf 13}  \\
\hline
\hline
CI  & 0.98 & 0.97 & 0.92 & 0.96 &  0.88  & 0.84   &  0.92 &  0.90 \\
$R_D$ & 1  &  1& 1 & 1  & 0.86  & 0.55 & 1 & 1  \\
\hline
\hline
{\bf Ref.} & \cite{hocking1976biometrics} & \cite{Spiegelhalter02} &    \cite{schwarz1978estimating} &   \cite{Hannan79}    &   \cite{AEDpaperNuestro}    &  \cite{SICpaper}  &  \cite{SICpaper} &  here \\ 
\hline
\end{tabular}
\end{center}
\end{table}

The first line represents the methods, the second line gives the number of suggested variables, and the last line contains the corresponding references. In \cite[Section 4-C]{OurPaperSound}, the results of that exhaustive analysis suggest that there are 7 very relevant variables  (level 1 of  \cite[Section 4-C]{OurPaperSound}),  other 7 relevant variables (level 2 of  \cite[Section 4-C]{OurPaperSound}) and other 2 variables in a level 3 of importance  \cite[Section 4-C]{OurPaperSound}, hence, totally 16 variables among very relevant, relevant and important ones. Note that the {suggested} numbers of SIC-95, SIC-99,\footnote{The numbers 95 and 99 are associated {with} the sum of some normalized weights (0.95 and 0.99) built by SIC which delivers  a  degree of confidence in the decision, as a confidence measure \cite{SICpaper}: closer to 100, the decision has a greater degree of assurance; {see also Appendix \ref{APP2}.}} BIC, the UAED and the ENV index are in line with this exhaustive analysis. ENV seems to provide a good compromise between SIC-95 and, on the opposite side,  SIC-99 and BIC. The result {provided} by ENV is quite close to the UAED so that $k_e=11$ with {a} high degree of reliability of $R_D=0.86$. 
\newline
Note that, in this experiment, the confidence measures provided by SIC {show} that SIC is more optimistic  {suggesting} of {a} more parsimonious model: for instance, the choice $k_e=7$ is fostered by SIC with $95\%$ of reliability, whereas $\mbox{CI}=84\%$ and $R_D=55 \%$. {Specifically,} the $R_D$ index warns that with $k_e=7$ we are missing $45\%$ of the effective number of variables (the ENV index is 12.74). 




\subsection{Variable selection in a biomedical classification problem with real data}\label{VSwithOscar}
In \cite{OscarPaper}, a feature selection analysis has been performed in order to find the most important variable for predicting patients at risk of developing nonalcoholic fatty liver disease among $35$ possible features. The authors have collected data from 1525 patients who attended the Cardiovascular Risk Unit of Mostoles University Hospital (Madrid, Spain) from 2005 to 2021, and {used} a random forest (RF) technique {as} a classifier  {to yield} a ranking of the components of  input vectors (of dimension $35$). They found that $4$ features were the most relevant according to the ranking and considering the medical experts' opinions: (a) insulin resistance, (b) ferritin, (c) serum levels of insulin, and (d) triglycerides. 
 \newline
{Here, we consider  
$$
V(k)=1-\mbox{accuracy-in-class}(k),
$$
 as a figure of merit (where we set $V(0)=0.5$, which represents a completely random binary classification).} The curve is obtained after ranking the 35 features \cite{OscarPaper}-\cite[Section 5.5]{SICpaper}. Note that with this choice of $V(k)$ we cannot apply BIC, AIC, and other standard information criteria, as shown in Table \ref{TablaIC}. However, the UAED \cite{AEDpaperNuestro}, SIC \cite{SICpaper}, and ENV can be applied. The results are given in the Table \ref{Table2}.

\begin{table}[!h]
\caption{Results in the variable selection example, in a classification problem with real data. } \label{Table2}
\begin{center}
\begin{tabular}{|c|c|c|c|c|c|}
\hline
 {\bf Scheme} &UAED &  SIC-90 & SIC-95 & SIC-99  &{\bf ENV} \\
\hline
$k_e$ & 3 & 2 & 3 & 9 &  {\bf 4}  \\
\hline
\hline
CI & 0.91 & 0.88 &0.91 & 0.98 & 0.92 \\
$R_D$ & 0.84 & 0.56 &0.84 & 1 & 1 \\
\hline
\end{tabular}
\end{center}
\end{table}

ENV gives a value of 3.5851, then suggests $k^*=4$ as the experts in  \cite{OscarPaper}. the UAED and SIC also provide results in line with this solution. In this experiment, in terms of confidence, the values of the confidence measures provided by SIC and the measures proposed here are very close: for the decision $k_e=3$ SIC provides {as} a confidence measure of 0.95, and $\mbox{CI}=0.91$, $R_D=0.84$ that are in the same line (high values similar to 0.95). The same observation can be done for $k_e=2, 4$, and $9$ ({except} the $R_D$ value - $0.56$ - for $k_e=2$, {which} alerts of a less confident decision missing almost 50$\%$ of the effective number of variables).

\section{Conclusions}\label{COsect}
An index of {an} effective number of variables has been proposed which has been inspired by the concept of maximum AUC in ROC curves and the Gini index.  
The introduced ENV index removes a dependence that we can find in the elbow detectors designed in the literature, i.e., the dependence on the maximum number of components $K$ analyzed.  We also introduce different measures of  uncertainty and reliability of the proposed solution (related to the ENV index). These novel confidence measures can be employed also jointly with the use {of} different information criteria such as the well-known AIC and BIC (where they can be applied, as we have shown in Section \ref{Ex2Sound}). {More generally, it is important to remark that the results obtained by the ENV index can be associated not only {with} an information criterion, but also {with} any other model selection procedure. Indeed, for computing CI we just need the knowledge  of the curve $V(k)$ and, for $R_D$, we need to use the ENV index $I_{\mathrm{ENV}}$ and a decision $k_e$, {which} can be obtained by any model selection scheme.}
Several comparisons with classical and recent schemes are provided in different experiments involving real datasets: we analyze two variable selection {problems}, one of them regarding a regression analysis and the other one involving a biomedical classification study. Related Matlab code is provided.

{\small
\section*{{\small Acknowledgment}}

The work was partially supported by the Young Researchers R\&D Project,  ref. num. F861 (AUTO-BA-GRAPH) funded by the Community of Madrid and Rey Juan Carlos University, by Agencia Estatal de Investigaci{\'o}n AEI  with project SP-GRAPH, ref. num. PID2019-105032GB-I00, with project POLI-GRAPH - Grant PID2022-136887NB-I00 funded by MCIN/AEI/10.13039/501100011033, and by Programa de Excelencia-Convenio Plurianual entre Comunidad de Madrid y la Universidad Rey Juan Carlos (ref. num. Y158/DF007003/30-06-2020, ref. num F840).
}

%
%

\bibliographystyle{IEEEtran}
\bibliography{bibliografia.bib}

\appendix

{
\section{Further considerations}\label{APP1}

In this appendix, we describe different ideal cases when $V(k)$ reaches the zero value at a specific $k$.
\newline
\newline
{\bf Scenario $V(0)\neq0$ and $V(1)=0$:} If we have a zero at $k=1$, i.e. $V(1)=0$, by assumption we have $V(1)=V(2)=...V(K)=0$, as shown in Figure \ref{Fig2Luca4}. In this case, computing Eq. \eqref{mostImp} or \eqref{ENVindexEq2}, we obtain $I_\texttt{ENV}= 1$. 
\newline
\newline
{\bf Scenario $V(0)\neq0$, $V(1)\neq0$ and $V(2)=0$:}
Hence,  by assumption we have $V(2)=V(3)=...V(K)=0$. Generally, we will have $1< I_\texttt{ENV}\leq 2$. The value of $I_\texttt{ENV}$ depends on the type of curve $V(k)$.
 The equality $I_\texttt{ENV}= 2$ is given if the decay is linear, i.e., $V(k)$ is a straight line in $0\leq k \leq 2$ (see the concept of  variable importance in Section \ref{SectReliab}).
\newline
\newline
{\bf Scenario $V(0)\neq0$, $V(1)\neq0$, $V(2)\neq0$...$V(k-1)\neq0$, until $V(k^*)=0$:} Hence,  by assumption we have $V(k^*)=V(k^*+1)=...V(K)=0$ as depicted in Figure \ref{Fig2Luca6}.
Generally, we will have $1< I_\texttt{ENV}\leq k^*$. Again, the value of $I_\texttt{ENV}$ depending on the type of curve $V(k)$: the equality $I_\texttt{ENV}= k^*$ is given if the decay is linear, i.e., $V(k)$ is a straight line in $0\leq k \leq k^*$ (see the concept of variable importance in Section \ref{SectReliab}).
\newline
 Recalling also the properties in Section \ref{FC_SECT} and due to the discrete nature of $k$, we cannot have value $0< I_\texttt{ENV}< 1$, but we can have $I_\texttt{ENV}= 0$, $I_\texttt{ENV}= 1$, and any value in the intervals $j-1\leq  I_\texttt{ENV}\leq j$ for all $j=2,...,K$. In summary, the values that the ENV index are the following $I_\texttt{ENV}\in\{0,[1,K]\}$. This makes sense since there is even a small decrease between $k=0$ and $k=1$ in $V(k)$, this means at least one component deserves to be considered in the model.

\section{Confidence measure by SIC}\label{APP2}

 A first attempt to provide a confidence measure has been given in  
the recent proposed spectral information criterion (SIC)  \cite{SICpaper} which returns:
\begin{itemize}
\item[{\bf (a)}] a suggestion regarding the position of the elbow $k_e$ (as the other information criteria);
\item[{\bf (b)}] but also provides a degree of certainty in the decision (as a confidence measure).
\end{itemize}
SIC associates to its decision with a number that is a sum of some normalized weights (e.g., often 0.95 and 0.99):  closer to 1  (100$\%$), the decision is more safe, meaning that we are confident in discarding the $K-k_e$ variables/components in the construction of the model.

 }

\end{document}